\documentclass[prb,twocolumn,amsmath,amssymb,superscriptaddress,aps,10pt]{revtex4}

\usepackage{graphicx}% Include figure files
\usepackage{dcolumn}% Align table columns on decimal point
\usepackage{bm}% bold math
\usepackage{gensymb}
\usepackage{amsmath}
\usepackage{color}
\usepackage{epsfig}
\usepackage{epstopdf}
\usepackage{soul}
\usepackage{booktabs} 
\usepackage{multirow}
\usepackage{microtype}
\usepackage{upgreek}
\usepackage{float}
\usepackage{fixltx2e}

\usepackage{hyperref}
\usepackage[utf8]{inputenc}
\usepackage[english]{babel}

\usepackage{xcolor}

\usepackage{tikz,xcolor,hyperref}

\definecolor{lime}{HTML}{A6CE39}
\DeclareRobustCommand{\orcidicon}{%
	\begin{tikzpicture}
	\draw[lime, fill=lime] (0,0) 
	circle [radius=0.16] 
	node[white] {{\fontfamily{qag}\selectfont \tiny ID}};
	\draw[white, fill=white] (-0.0625,0.095) 
	circle [radius=0.007];
	\end{tikzpicture}
	\hspace{-2mm}
}

\foreach \x in {A, ..., Z}{%
	\expandafter\xdef\csname orcid\x\endcsname{\noexpand\href{https://orcid.org/\csname orcidauthor\x\endcsname}{\noexpand\orcidicon}}
}
\begin{document}

%===================
% \preprint{APS/123-QED}
%===================

\title{Dynamic Terahertz Beamforming Based on \\ Magnetically Switchable Hyperbolic Materials} 

\newcommand{\orcidauthorA}{0000-0001-6178-0061} % william
\newcommand{\orcidauthorB}{0000-0003-1742-9957} % jorge
\newcommand{\orcidauthorC}{0000-0001-5698-6639} % Danilo
\newcommand{\orcidauthorD}{0000-0002-2095-7008} % Edwin

\author{William O. F. Carvalho\orcidA{}}
\email{williamofcarvalho@gmail.com}
\affiliation{Federal University of Itajub\'a (UNIFEI), 37500-903 Itajub\'a, MG, Brazil.}
\author{E. Moncada-Villa\orcidD{}}
\affiliation{Escuela de F\'isica, Universidad Pedag\'ogica y Tecnol\'ogica de Colombia, Avenida Central del Norte 39-115, Tunja, Colombia}
\author{J. R. Mej\'ia-Salazar\orcidB{}}
\affiliation{National Institute of Telecommunications (Inatel), Santa Rita do Sapuca\'i, MG, 37540-000, Brazil}
\author{Danilo H. Spadoti\orcidC{}}
\affiliation{Federal University of Itajub\'a (UNIFEI), 37500-903 Itajub\'a, MG, Brazil.}

\date{\today}% It is always \today, today,

\begin{abstract}
In this work, we introduce a concept to enable dynamic beamforming of terahertz (THz) wavefronts using applied magnetic fields ($\textbf{B}$). The proposed system exploits the magnetically switchable hyperbolic dispersion of the InSb semiconductor. This phenomenology, combined with diffractive surfaces and magnetic tilting of scattered fields, allows the design of a metasurface that works with either circularly or linearly polarized wavefronts. In particular, we demonstrate numerically that the transmitted beam tilting can be manipulated with the direction and magnitude of $\textbf{B}$. Numerical results, obtained through the finite element method (FEM), are qualitatively supported by semi-analytical results from the generalized dipole theory. Motivated by potential applications in future Tera-WiFi active links, a proof of concept is conducted for the working frequency $f=300$~GHz. The results indicate that the transmitted field can be actively tuned to point in five different directions with beamforming of $\pm45^{\circ}$, depending on the magnitude and direction of $\textbf{B}$. In addition to magnetic beamforming, we also demonstrate that our proposal exhibits magnetic circular dichroism (MCD), which can also find applications in magnetically tunable THz isolators for one-way transmission/reflection.
\end{abstract}

\maketitle

%\tableofcontents

%====================================
\section{Introduction}

The electromagnetic gap between electronic and optical regimes, widely-known as the terahertz (THz) band (from $0.1$~THz to $10$~THz), has been drawing increasing attention during the last years. Such interest is mainly driven by a plethora of applications, including spectroscopy,\cite{fu2022applications} imaging,\cite{jiang2022machine} security,\cite{jornet2023wireless} sensing,~\cite{Wang2023} and communications,~\cite{luo2019graphene,pant2023thz,wang2021recent,xu2021graphene,chen2022terahertz,Li2022,tan2023terahertz,you2023mechanically} among others. The first challenge the researchers tackled was the development of THz sources and detectors, which today are consolidated into a number of different approaches and commercially available devices.\cite{Davies2004} Nevertheless, these THz equipments are passive and therefore exhibit limited functionalities. For example, the need to use highly directional beams in THz wireless broadcasting (generated from high-gain antennas designed to surpass the large free-space path loss~\cite{Song2022}) restricts communication to two fixed points, called the transmit and receive antennas. Thus, researchers are currently focused on finding mechanisms/alternatives that allow active manipulation of THz wavefronts. A promising approach to achieve the latter goal is the use of metasurfaces~\cite{Yu2014} (i.e. artificial materials comprising two-dimensional arrays of subwavelength ``meta-atoms'') engineered to dictate (at will) reflectance, transmittance, and absorbance features.~\cite{Yu2011} Although metasurfaces operating in the microwave and optical regimes can be actively manipulated using diodes (e.g. varactors)~\cite{Taravati2022,Zhang2022} and tunable light-matter interactions,~\cite{Neshev2018,Du2022,Cortes2022} respectively, innovative approaches are still needed for the THz range. Some attempts include the use of liquid crystals~\cite{Wu2020,Fu2022,Liu2022,Zhuang2023} and phase-change materials.~\cite{Wang2020,Chen2022,Zeng2023} However, these latest platforms require precise control of the temperature of the building materials, hindering large-scale implementation for terahertz systems. Moreover, temperature changes conventionally occur at low speeds, thus limiting the velocity of operation. Hence, the search for fast and dynamic beamforming of THz wavefronts with large-scale implementable solutions remains an open problem. 

Inspired by the use of magnetic fields to manipulate the optical radiated beams from magnetoplasmonic nanoantennas,\cite{Damasceno2023,Carvalho2023,Carvalho2023a} we hypothesized that magnetic fields can also be employed for dynamic THz beamforming through magnetically tunable metasurfaces. In the THz range, Faraday and polar magneto-optical (MO) Kerr effects (PMOKE) were recently used for isolators and filters. These achievements have been made using MO metasurfaces comprising the InSb material,~\cite{Lin2018,li2020terahertz,fan2021magnetically,Tan2021} whose conduction electrons are strongly affected by applied magnetic fields. Nevertheless, Faraday and PMOKE phenomena only induce polarization rotation effects, which do not tilt the transmitted/radiated beams, as demonstrated analytically and numerically in Ref.~[\!\!\citenum{Carvalho2023a}]. Moreover, it is well-known from the first observations of MO phenomena by Faraday~\cite{Faraday1846} and Kerr~\cite{Kerr1877,Kerr1878} that the Faraday and PMOKE effects are at least one order of magnitude higher than their in-plane counterparts, i.e., the longitudinal and transverse MO Kerr effects (LMOKE and TMOKE). Therefore, configurations different to Faraday and PMOKE are conventionally studied in the visible and infrared wavelength ranges, where strong light-matter interactions are used to enhance their amplitudes.~\cite{Rizal2021}

\begin{figure}[!t]
\centerline{\includegraphics[width=0.9\columnwidth]{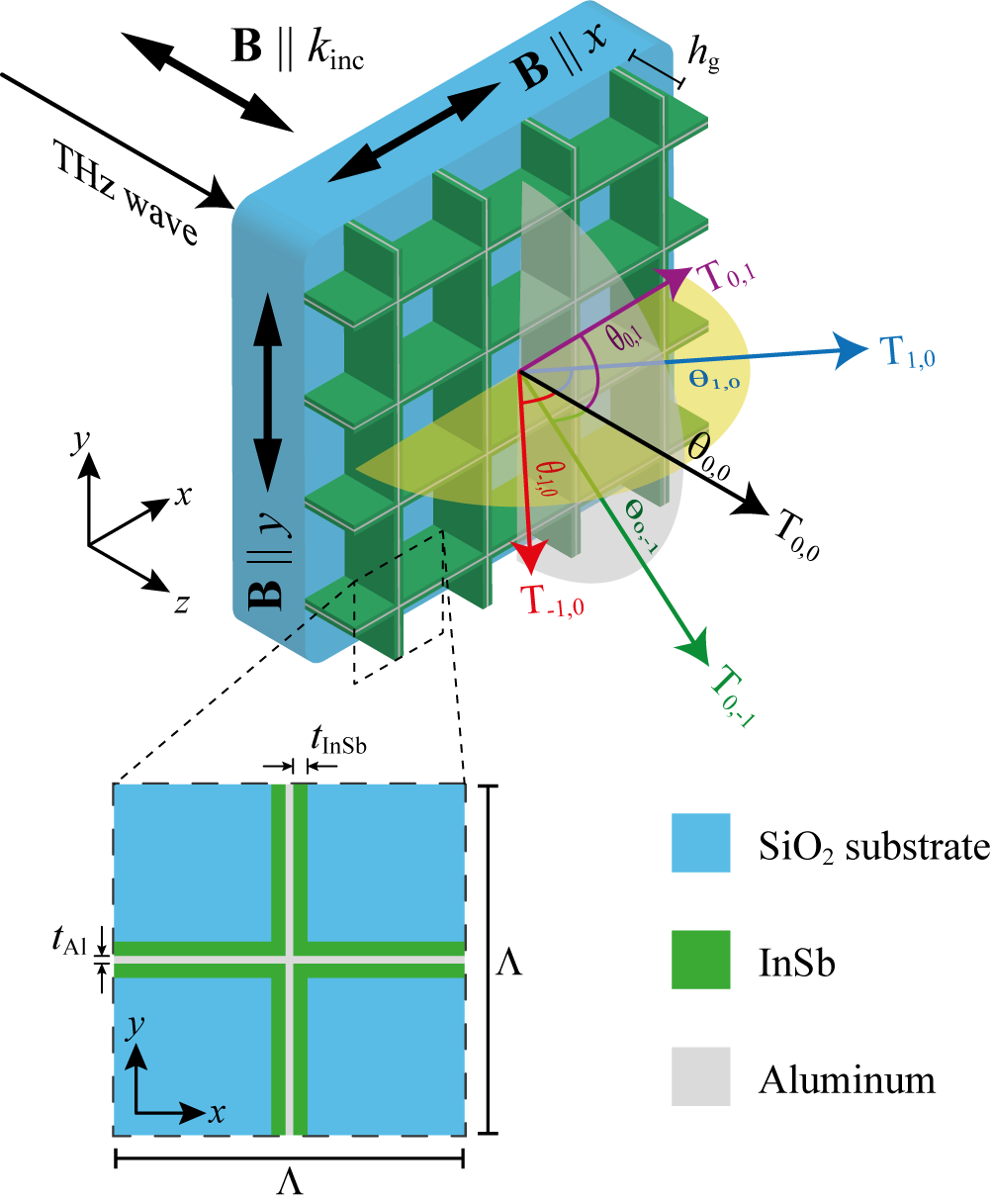}}
\caption{Schematic of the MO metasurface. The inset shows the material composition and geometrical parameters of the unit cell.}
\label{fig1}
\end{figure}

In this work, we demonstrate a magnetically active metasurface that allows the dynamic beamforming of THz wavefronts. Our proposal consists of a fishnet-like metasurface made of aluminum (Al) and the indium antimonide (InSb) semiconducting material, as illustrated in Figure~\ref{fig1}. The inset of this figure shows a cross of Al material symmetrically covered by a thin film of InSb on each side. The Al is a metallic material conventionally used in the fabrication of resonant THz metasurfaces. Indeed, the cross-like design of Al elements can be easily found in recent works.~\cite{Kenney2017} Hence, we focused here on exploiting the MO activity of the InSb material (which can be deposited over the entire surface and then molded using ion-milling and focused ion beam (FIB) techniques~\cite{Jany2015} for the active manipulation of THz wavefronts. In particular, the InSb material exhibits a magnetically switchable hyperbolic dispersion, i.e., the regular to hyperbolic dispersion can be activated/deactivated using a static magnetic field.\cite{Li2012} The later feature, in combination with geometric design, is used to allow the excitation of asymmetric dipole resonances when the direction of the in-plane magnetic field is flipped. These asymmetric resonances dictate the preferred diffraction order ($\pm1$) for the transmitted wavefronts, as qualitatively demonstrated using the dipole theory for the unit cell.

%========FIGURE 1============
%==========================

%====================================
\section{Methodology}

A metasurface composed of a two-dimensional periodic arrange of metallic crosses (made of Al) is considered, as depicted in the inset of Figure~\ref{fig1}. The period length ($\Lambda$) and thickness ($t_{\rm Al}$) of the crosses are selected to resonate at the THz frequency range. Since the frequency $f=300$~GHz has applicability for future wireless communications,\cite{IEEE802.15.3d-2017,petrov2020ieee} we used it as the working frequency to prove our concept. MO activity is introduced in the system by using uniform and symmetrically placed slabs of InSb material on each side of the metal building blocks (see the inset of Figure~\ref{fig1}). The InSb slabs have a thickness $t_{\rm InSb}$ and the same height of the Al crosses, i.e., $h_{\rm g}$ according to Figure~\ref{fig1}. The structure, which can be fabricated by a combination of different techniques,~\cite{Kenney2017,Jany2015} is considered grown on a conventional SiO$_2$ substrate. Under the effect of an externally applied magnetic field, the off-diagonal components of the InSb material have non-zero values. For the magnetic field ($\textbf{B}$) pointing along one of the main axis, we have
%==================================
\begin{equation}\label{eqtensor1}
\tilde{\varepsilon}_{\rm InSb}
= 
\begin{pmatrix}
\varepsilon_{\rm \parallel}       & 0                           & 0                          \\
0                             & \varepsilon_{\rm \perp}         & i\varepsilon_{\rm off} \\
0                             & -i\varepsilon_{\rm off}         & \varepsilon_{\rm \perp}
\end{pmatrix},
\end{equation}
%==================================
for $\textbf{B}$ parallel to the $x$-axis, 
%==================================
\begin{equation}\label{eqtensor2}
\tilde{\varepsilon}_{\rm InSb}
= 
\begin{pmatrix}
\varepsilon_{\rm \perp}       & 0                           & i\varepsilon_{\rm off}     \\
0                             & \varepsilon_{\rm \parallel} & 0                           \\
-i\varepsilon_{\rm off}       & 0                           & \varepsilon_{\rm \perp}
\end{pmatrix},
\end{equation}
%==================================
for $\textbf{B}$ along the $y$-axis and 
%==================================
\begin{equation}\label{eqtensor3}
\tilde{\varepsilon}_{\rm InSb}
= 
\begin{pmatrix}
\varepsilon_{\rm \perp}       & i\varepsilon_{\rm off}      & 0                           \\
-i\varepsilon_{\rm off}       & \varepsilon_{\rm \perp}     & 0                           \\
0                             & 0                           & \varepsilon_{\rm \parallel}
\end{pmatrix},
\end{equation}
%==================================
for $\textbf{B}$ applied along the $z$-axis. The permittivities $\varepsilon_{\rm \perp}$, $\varepsilon_{\rm \parallel}$, and $\varepsilon_{\rm off}$ are described by
%==================================
\begin{align}\label{eqepsperp}
\varepsilon_{\rm \perp} &= \varepsilon_{\infty} - \frac{\omega_{\rm p}^{2}\left(\omega+i\gamma\right)}{\omega\left[(\omega+i\gamma)^{2}-\omega_{\rm c}^{2}\right]}, \\
\varepsilon_{\rm \parallel} &= \varepsilon_{\infty} - \frac{\omega_{\rm p}^{2}}{\omega(\omega+i\gamma)}, \\
\varepsilon_{\rm off} &= -\frac{\omega_{\rm p}^{2}\omega_{\rm c}}{\omega\left[(\omega+i\gamma)^{2}-\omega_{\rm c}^{2}\right]},
\end{align}
%==================================
where the subscripts $\parallel$ and $\perp$ are used to indicate the permittivity components that are parallel and perpendicular to $\textbf{B}$, respectively. $\varepsilon_{\infty}$ is the high frequency permittivity and $\omega_{\rm p}=\sqrt{N e^2 / \varepsilon_{0} m^{*}}$ is the plasma frequency. $N$ represents the carrier density, $e$ the electron charge, $\varepsilon_0$ the vacuum permittivity and $m^*=0.014m$ the electron effective mass for the InSb material. $\omega$ is the angular frequency, $\gamma=e/\mu m^{*}$ is the damping constant, $\mu$ is the carrier mobility, and $\omega_{\rm c} = e B / m^{*}$ ($B=\left|\textbf{B}\right|$) is the cyclotron frequency. Since $N$ is strongly temperature-dependent, we employed an empirical equation from the available literature $N=5.76\times 10^{14} \times T_{\rm K}^{1.5} \exp{\left(\frac{-1.5\times 10^{3}}{T_{\rm K}}\right)}$,\cite{tan2019nonreciprocal} where $T_{\rm K}$ is the temperature in Kelvin. The permittivities (considered isotropic) of Al and SiO$_2$ were taken from the available experimental literature.\cite{hagemann1975absorption,naftaly2007terahertz} 

The grating features of the proposed metasurface provide different orders and efficiencies for diffractively transmitted fields ($T_{m,n}$), as depicted by arrows in Figure~\ref{fig1}. As we are using a symmetric two-dimensional grating ($\Lambda_x = \Lambda_y = \Lambda$), the corresponding phase-matching condition for normal incidence is expressed by\cite{zhou2021polarization}
%==================================
\begin{equation}\label{eqgrating}
    m^2 + n^2 = \frac{\Lambda^2\sin^2{\theta_{m,n}}}{\lambda^2}
\end{equation}
%==================================
where $m$ and $n$ are integer numbers representing the diffraction orders, with diffracted angle $\theta_{m,n}$. Since we are showing a proof-of-concept for future wireless THz broadcasting, we focused only (for simplicity) on transmitted diffraction orders with $\theta_{m,n}=45^{\circ}$ for $(m,n)=(0,\pm1)$ and $(m,n)=(\pm1,0)$. Thus, we set $\Lambda~=~1414~\upmu$m, according to Eq.~\eqref{eqgrating}, to allow only orders $0^{\rm th}$ and $1^{\rm st}$ to be transmitted.

Full-wave numerical simulations were carried out using the finite element method (FEM), within the commercial software COMSOL Multiphysics\textsuperscript{\textregistered}. Refined irregular-mesh-sizes were used for improved precision, with finer meshes near the boundaries of the grating. Floquet periodic boundary conditions were used along the $x$- and $y$-axes, whereas absorbing perfectly matched layers (PMLs) were considered along the $z$-boundaries. All results were obtained for normal incident THz fields, with a frequency $f=300$~GHz.

In addition to full-wave numerical simulations, semi-analytical results were obtained within the discrete dipole approximation (DDA)\cite{Novotny2012}, where we consider a system of $N$ MO point dipoles, at arbitrary positions $\mathbf r_i$, having polarizabilities
\begin{equation}\label{pol}
\hat\alpha_i = \left(\hat{\mathbf 1} -i\frac{k^3}{6\pi}\hat{\alpha}_{i,0}\right)^{-1}\hat\alpha_{i,0},
\end{equation}
with $\hat\alpha_{i,0}=3v_i[\hat\epsilon_i-\hat{\mathbf 1}][\hat\epsilon_i+2\hat{\mathbf 1}]^{-1}$ for the static (non-radiative) polarizability, where $v_i$ and $\hat\epsilon_i$ are the volume and dielectric permittivity tensor of each dipole, respectively. Under the excitation of a monochromatic electromagnetic plane wave, the scattered electric and magnetic fields (at a position $\mathbf r$) can be expressed as\cite{Novotny2012,Maccaferri2016,Ott2018}
\begin{eqnarray}\label{Efield}
	\mathbf E(\mathbf r) &=& \frac{k^2}{\varepsilon_0}\bar{ G}^{\rm E}\cdot \bar{\mathbf p} ,\\  \label{Hfield}
	\mathbf H(\mathbf r) &=& \frac{k^2}{\varepsilon_0} \bar{ G}^{\rm H}\cdot \bar{\mathbf p},
\end{eqnarray}
where $k=\omega/c$ is wave vector magnitude of the emitted radiation with angular frequency $\omega$ ($c$ is the speed of light in vacuum). $\bar{ G}^{\rm \beta}=(\hat{ G}^{\rm \beta} (\mathbf r,\mathbf r_1),\hat{ G}^{\rm \beta} (\mathbf r,\mathbf r_2),...,\hat{ G}^{\rm \beta} (\mathbf r,\mathbf r_N))$ (with $\beta$ representing E or H) are $3\times 3N$ matrices built with the following electric and magnetic dyadic Green's functions
\begin{widetext}
\begin{eqnarray}\label{green}
 \hat{ G}^{\rm E}(\mathbf r,\mathbf r_i) & = & \frac{e^{ikR_i}}{4\pi R_i } \left[ \left(1+\frac{ikR_i-1}{(kR_i)^2}\right)\hat {\mathbf 1} \right. + \left. \left( \frac{3-3ikR_i-(kR_i)^2}{(kR_i)^2}\right) \hat{\mathbf R}_i\otimes \hat{\mathbf R}_i\right] \\
 \hat{ G}^{\rm H}(\mathbf r,\mathbf r_i) & = & \frac{e^{ikR_i}}{4\pi Z_0R_i} \left(1+\frac{i}{kR_i}\right) 
\times \left( \begin{array}{ccc} 
0  			  & 		z-z_i  				   &-(y-y_i)\\
-(z-z_i)							   &	 0    &     x-x_i   \\
y-y_i   & -(x-x_i)						& 0
\end{array} \right),
 \end{eqnarray}
 \end{widetext}
where $Z_0=\sqrt{\mu_0/\varepsilon_0}$ is the vacuum electromagnetic impedance, $\mathbf R_i=\mathbf r - \mathbf r_i$, $R_i=|\mathbf{R}_i|$, and $\hat {\mathbf{R}}_i=\mathbf{R}_i/R_i$. 

The $3N\times 1$ supervector $\bar{\mathbf p}=(\mathbf p_1, \mathbf p_2,...,\mathbf p_N)^{\rm T} $, in Equations \eqref{Efield}  and \eqref{Hfield}, contains the dipolar moments $\mathbf p_i=\varepsilon_0\hat\alpha_i\mathbf E_{{\rm{exc}},i}$ of the system. Then, we can write
\begin{equation}\label{dipolar}
	\bar{\mathbf p} =\varepsilon_0\bar\alpha \bar{\mathbf E}_{\rm{exc}},
\end{equation}
where $\bar\alpha={\rm diag}(\hat\alpha_1,\hat\alpha_2,...,\hat\alpha_N)$ and $\bar{\mathbf E}_{\rm{exc}}=(\mathbf{E}_{{\rm{exc}},1},\mathbf{E}_{{\rm{exc}},2},...,\mathbf{E}_{{\rm{exc}},N})^{\rm T}$ is a $3N\times1$ supervector comprising the fields that excite all the dipoles in the system. These exciting fields can be determined by a set of $N$ equations of the form \cite{ekeroth2017}
\begin{equation}\label{DDA1}
	\mathbf{E}_{{\rm{exc}},i} = \mathbf{E}_{{\rm{inc}},i}+ k_0^2\sum_{j\neq i}\hat G_{ji}\hat \alpha_{j}\mathbf{E}_{{\rm{exc}},i},
\end{equation}
where $\mathbf{E}_{{\rm{inc}},i}=\mathbf E_0e^{i(\mathbf k\cdot\mathbf r_i-\omega t)}$ is the incident field upon the $i-$th dipole, and $\hat G_{ji}=\hat G_{ji}(\mathbf r_j,\mathbf r_i)$ is the inter-dipolar Green's function, given by
\begin{eqnarray}\label{greendip}
 \hat{ G}(\mathbf r_j,\mathbf r_i) & = & \frac{e^{ikR_{ji}}}{4\pi R_{ji}} \left[ \left(1+\frac{ikR_{ji}-1}{(kR_{ji})^2}\right)\hat {\mathbf 1} \right. 
 \\\nonumber && + \left. \left( \frac{3-3ikR_{ji}-(kR_{ji})^2}{(kR_{ji})^2}\right) \hat{\mathbf R}_{ji}\otimes \hat{\mathbf R}_{ji}\right], 
 \end{eqnarray}
with $\mathbf R_{ji}=\mathbf r_j-\mathbf r_i$. The latter system of equations can be expressed compactly as 
\begin{equation}\label{field_exc}
	\bar{\mathbf E}_{\rm{exc}}=\bar {T} \bar{\mathbf E}_{\rm{inc}},
\end{equation}
where $\bar T=(\hat{\mathbf{1}}-k_0^2\Delta\bar G\bar\alpha )^{-1}$ and $\Delta\bar G=\bar G-{\rm diag}\{\bar G\}$ are $3N\times3N$ matrices, with $[\bar G]_{ij}=\hat G_{ij}$. Therefore, using the Equations \eqref{dipolar} and \eqref{field_exc}, we can rewrite the scattered electric and magnetic fields as
\begin{eqnarray}\label{Efield2}
	\mathbf E(\mathbf r) &=& k^2\bar{ G}^{\rm E}\bar\alpha\bar T\bar{\mathbf E}_{{\rm inc}} ,\\  \label{Hfield2}
	\mathbf H(\mathbf r) &=&  k^2\bar{ G}^{\rm H}\bar\alpha\bar T\bar{\mathbf E}_{{\rm inc}},
\end{eqnarray}
from where the scattered power per unit area $\mathbf S(\mathbf r)=\frac{1}{2}{\rm Re}\{\mathbf E(\mathbf r)\times \mathbf H^*(\mathbf r) \}$ can be calculated.

In the case of two identical dipoles, as it is being considered in this work (for simplicity), the expressions above can be reduced to
\begin{eqnarray}
	\mathbf E(\mathbf r) &=&k^2\hat{ G}^{\rm E}(\mathbf r, \mathbf r_1)\hat\alpha[  \hat T_{11}\mathbf E_{\rm inc}(\mathbf r_1) +  \hat T_{12}\mathbf E_{\rm inc}(\mathbf r_2)   ]\\\nonumber
	&+&k^2\hat{ G}^{\rm E}(\mathbf r, \mathbf r_2)\hat\alpha[  \hat T_{21}\mathbf E_{\rm inc}(\mathbf r_1) +  \hat T_{22}\mathbf E_{\rm inc}(\mathbf r_2)   ],\\
	\mathbf H(\mathbf r) &=&k^2\hat{ G}^{\rm H}(\mathbf r, \mathbf r_1)\hat\alpha[  \hat T_{11}\mathbf E_{\rm inc}(\mathbf r_1) +  \hat T_{12}\mathbf E_{\rm inc}(\mathbf r_2)   ]\\\nonumber
	&+&k^2\hat{ G}^{\rm H}(\mathbf r, \mathbf r_2)\hat\alpha[  \hat T_{21}\mathbf E_{\rm inc}(\mathbf r_1) +  \hat T_{22}\mathbf E_{\rm inc}(\mathbf r_2)   ],
\end{eqnarray}
with
\begin{eqnarray}
	\hat T_{11}&=& (\hat{\mathbf 1} - k^4\hat G_{12}\hat\alpha\hat G_{12}\hat\alpha)^{-1},\\
	\hat T_{12}&=& k^2\hat T_{11}\hat G_{12}\hat\alpha,\\
	\hat T_{21}&=&k^2\hat G_{12}\hat\alpha\hat T_{11},\\
	\label{T22}
	\hat T_{22}&=&\hat{\mathbf 1} +k^4\hat G_{12}\hat\alpha\hat T_{11}\hat G_{12}\hat\alpha,
\end{eqnarray}
from where numerical results for the scattered energy were calculated.

%========================================================
\section{Results and Discussion}
%=====================================================

Let us first discuss the permittivity values for the InSb material at $f=300$~GHz and $T_{\rm K} = 200$~K, summarized in Table~\ref{table1}. For $B=0$~T we have an isotropic ($\varepsilon_{\parallel}=\varepsilon_{\perp}$ and $\varepsilon_{\rm off}=0$) metallic behavior, whilst for $B=\pm3$~T the InSb material behaves as a HMM,\cite{lobet2023new} i.e., $\varepsilon_{\rm \perp}\varepsilon_{\rm \parallel}<0$. In particular, the unique feature of having $\varepsilon_{\rm \perp}>0$ makes InSb slabs behave like high-refractive-index (HRI) dielectric media along those specific directions, which we exploit to produce magnetically tunable dipolar resonances. The latter produces an effect analogous to the case of nanoantennas,\cite{Carvalho2023a,Carvalho2023,Damasceno2023} but in each unit cell of the metasurface. Therefore, we must find the geometric parameters at which the collective effects, through the constructive phase interference, produce the maximum beamforming. These parameters were found (using the optimization technique in COMSOL) as $t_{\rm InSb}=58~\upmu$m, $t_{\rm Al}=30~\upmu$m, $h_{\rm g}=1000~\upmu$m and $\Lambda = 1414~\upmu$m, which are considered constant along the entire work.

%==========================================
\begin{table}[!b]
 \caption{Components of $\varepsilon_{\rm Al}$,\cite{hagemann1975absorption} $\varepsilon_{\rm SiO_2}$\cite{naftaly2007terahertz} and $\tilde{\varepsilon}_{\rm InSb}$\cite{tan2019nonreciprocal} at $f=300$~GHz and $T_{\rm K} = 200$~K, under the influence of $B$.}\label{table1}
 \resizebox{\columnwidth}{!}{
  \begin{tabular}{cc||cc}
    \hline
    Parameter                       & Value           & Parameter                         & Value\\
    \hline
$\varepsilon_{\rm Al}$              & $-8.08\textrm{e}5+i1.95\textrm{e}6$   & $\varepsilon_{\rm SiO_2}$         & $3.8+i0.02$\\
$\varepsilon_{\infty}$              & 15.68           & $\varepsilon_{\rm \parallel}$     & $-31.21+i20.73$\\
$\varepsilon_{\rm \perp}$, $B=0$~T & $-31.21+i20.73$ & $\varepsilon_{\rm off}$, $B=0$~T  & $0$\\
$\varepsilon_{\rm \perp}$, $B=+3$~T & $15.82+i0.06$   & $\varepsilon_{\rm off}$, $B=+3$~T & $-2.81-i0.006$\\
$\varepsilon_{\rm \perp}$, $B=-3$~T & $15.82+i0.06$   & $\varepsilon_{\rm off}$, $B=-3$~T & $2.81+i0.006$\\
    \hline
  \end{tabular}
}
\end{table}
%%=====================================================

%========FIGURE 2============
\begin{figure}[!t]
\centerline{\includegraphics[width=.9\columnwidth]{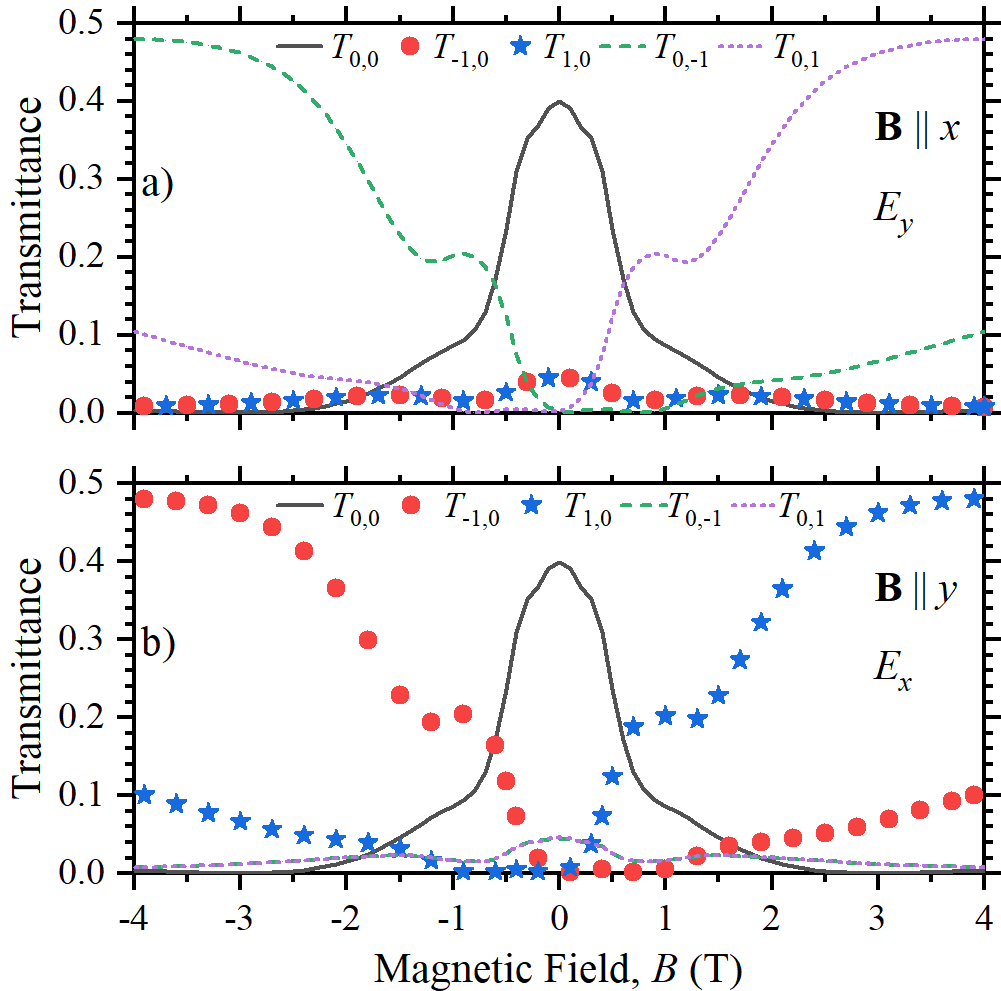}}
\caption{Diffractively transmitted orders for a LP THz wavefront, with $f=300$~GHz, as function of the externally applied magnetic field amplitude $B$. Results are comparatively shown for a) $\textbf{E}\parallel \hat{\textbf{y}}$ (with $\textbf{B} \parallel \hat{\textbf{x}}$) and b) $\textbf{E}\parallel \hat{\textbf{x}}$ (with $\textbf{B} \parallel \hat{\textbf{y}}$). Solid, dashed, and dotted lines are for $T_{0,0}$, $T_{0,-1}$, and $T_{0,1}$, whilst solid circles and stars are for $T_{-1,0}$ and $T_{1,0}$, respectively.}
\label{fig2}
\end{figure}
%==========================

%========FIGURE 3============
\begin{figure}[!b]
\centerline{\includegraphics[width=.9\columnwidth]{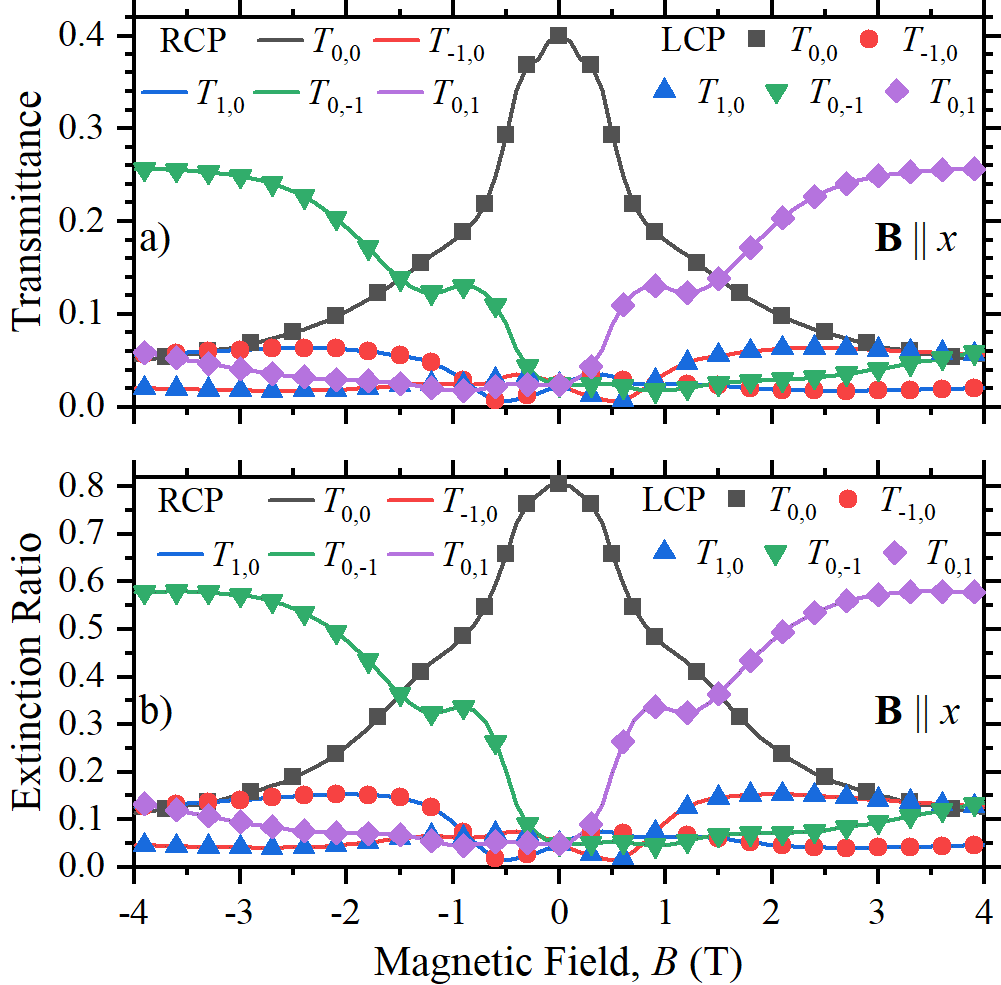}}
\caption{a) Diffractively transmitted orders and b) extinction ratio for a CP THz wavefront, with $f=300$~GHz, as function of the externally applied magnetic field amplitude $B$. Solid lines are for RCP, whereas solid symbols are for LCP wavefronts, associated to $T_{0,0}$ (black), $T_{-1,0}$ (red), $T_{1,0}$ (blue), $T_{0,-1}$ (green) and $T_{0,1}$ (purple).
}
\label{fig3}
\end{figure}
%==========================

%========FIGURE 4============
\begin{figure*}[!t]
\centerline{\includegraphics[width=1.8\columnwidth]{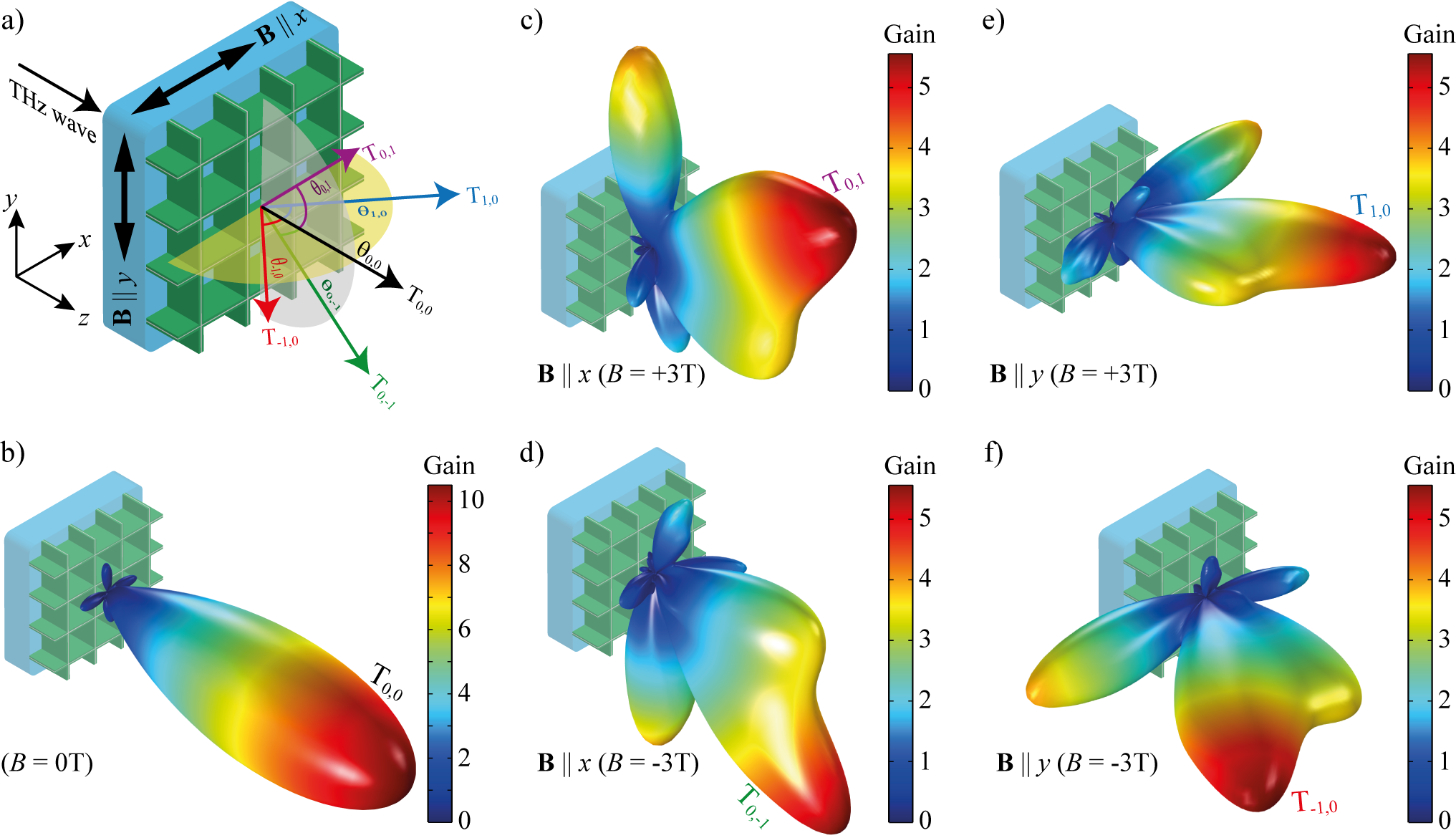}}
\caption{a) Metasurface and the corresponding diffracted beams $T_{n,m}$ are illustrated for eye-guide. The corresponding 3D radiation patterns are plotted for b) $B=0$ and $B=\pm3$~T along c)-d) $\pm\hat{x}$ and e)-f) $\pm\hat{y}$ directions.}
\label{fig4}
\end{figure*}
%==========================

Figure~\ref{fig2} shows the transmitted diffraction orders $T_{m,n}$, for a linearly polarized (LP) THz wave, as function of the externally applied magnetic field $B$. To illustrate the symmetry effects of the square unit cell, we calculate $T_{m,n}$ for two different electric field polarizations of the incident wave. First, numerical results for $T_{m,n}$ with $\textbf{E}\parallel \hat{\textbf{y}}$ and $\textbf{B} \parallel \hat{\textbf{x}}$ are shown in Figure~\ref{fig2}a). The sign of $B$ on the horizontal axis indicates the direction of $\textbf{B}$ along the magnetized axis. The transmission for $B=0$ is dominated by the diffraction orders ($m=n=0$) along the $z$-axis, as observed for $T_{0,0}$ in Figure~\ref{fig2}a). Interestingly, transmission becomes increasingly dominated by $T_{0,1}$ and $T_{0,-1}$ along the $yz$-plane as $\textbf{B}\neq\textbf{0} \parallel \hat{\textbf{x}}$ increases. Symmetric results are observed for the $xz$-plane when the applied magnetic field and electric field polarization are rotated by $90^{\circ}$, as noticed from Figure~\ref{fig2}b) for $\textbf{E}\parallel \hat{\textbf{x}}$ and $\textbf{B} \parallel \hat{\textbf{y}}$. This behavior is due to the HRI resonances associated with the HMM dispersion of the InSb material, as will be explained later for the more general case of circular polarization (CP).

%========FIGURE 5============
\begin{figure}[!b]
\centerline{\includegraphics[width=1\columnwidth]{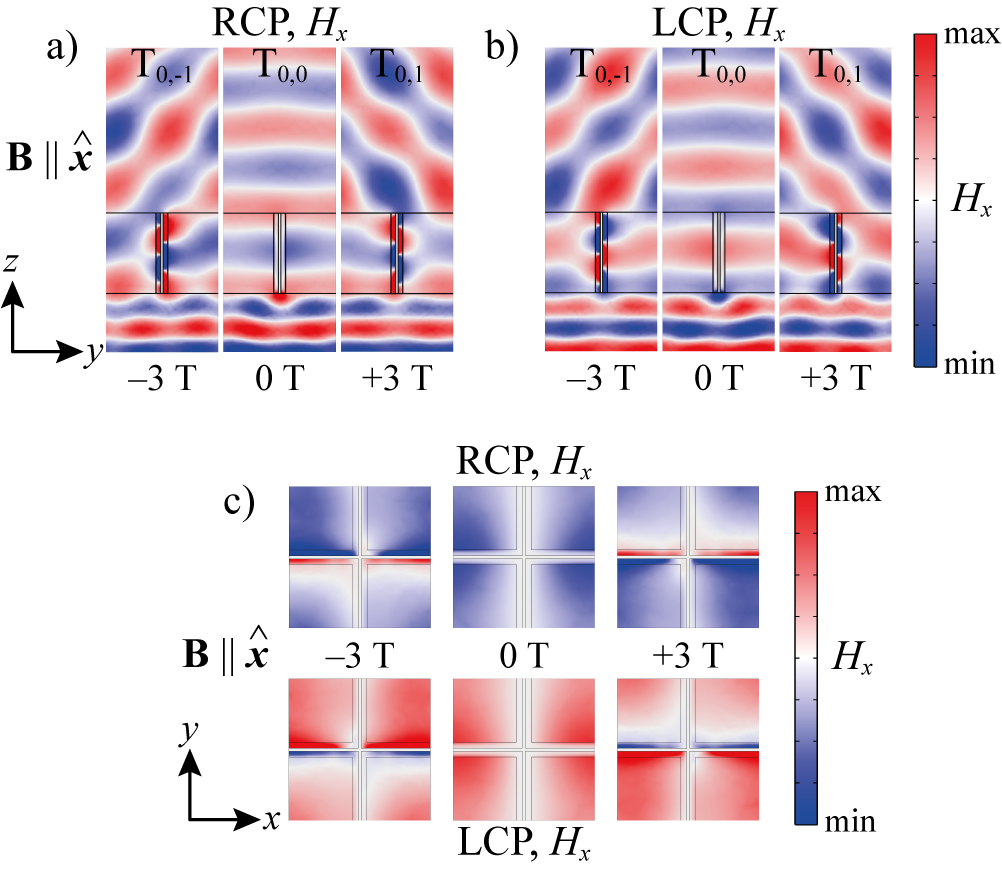}}
\caption{a)-b) Side and c) upper view of $H_x$ near-field profiles for a) RCP and b) LCP incident wavefronts with $B=(-3, 0, +3)$~T along $x$-axis.}
\label{fig5}
\end{figure}
%==========================

%========FIGURE 6============
\begin{figure*}[!t]
\centerline{\includegraphics[width=1.8\columnwidth]{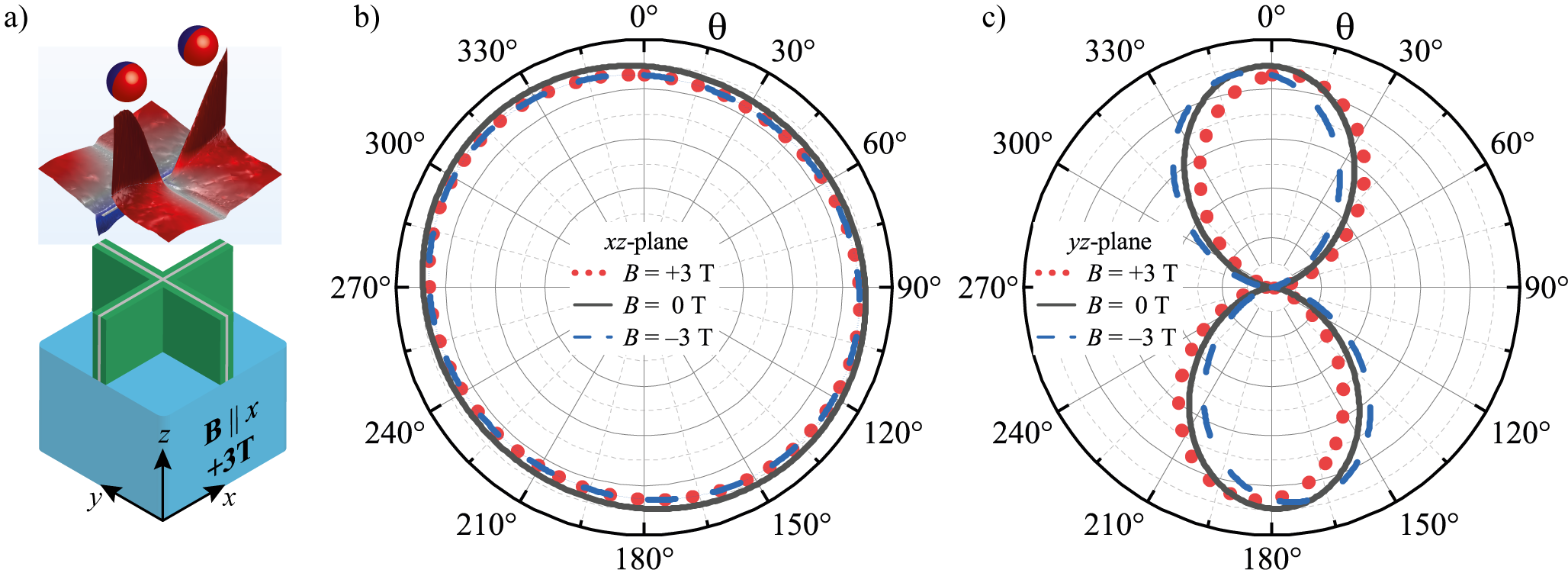}}
\caption{a) Unit cell and its representative dipolar behavior using LCP under $B=+3~\textrm{T} \parallel \hat{x}$. Scattered electromagnetic energy along the b) $xz$ and c) $yz$ planes, respectively, for $B=0$~T~(black), $B=+3$~T~(red) and $B=-3$~T~(blue).
}
\label{fig6}
\end{figure*}
%==========================

From here on, CP THz wavefronts are used for a more general discussion. The electric field component of CP waves is written as $\textbf{E}=(E_x\hat{\textbf{x}} \pm iE_y\hat{\textbf{y}})e^{-ik_{\rm inc}z}$, simultaneously having $E_{x}$ and $E_{y}$ components, where the sign $+/-$ denotes the right/left CP (RCP/LCP) state. In analogy to results in Figure~\ref{fig2}, we calculated the transmitted diffraction orders $T_{m,n}$ for the RCP and LCP configurations. Numerical results are shown in Figure~\ref{fig3}a) for the case of $\textbf{B} \parallel \hat{\textbf{x}}$, with magnetically induced tilting along the $yz$-plane. Although not shown here, calculations for $\textbf{B} \parallel \hat{\textbf{y}}$ exhibited the same symmetry properties discussed in Figure~\ref{fig2}b), i.e., transmitted fields tilted along the $xz$-plane are induced by the external magnetic field. For a quantitative comparison among different transmitted diffracted orders, we calculate the extinction ratio ($\textrm{ER} = T_{m,n}/T_{\rm total}$, where $T_{\rm total}$ is the sum of all transmitted modes) in Figure~\ref{fig3}b). From this last figure it can be seen that in the absence of magnetic field ($B=0$) $81\%$ of the transmitted power corresponds to the $T_{0,0}$ mode. In contrast, the transmittances $T_{0,\pm1}$ and $T_{\pm1,0}$ achieve values as high as 57\% of the total transmitted power (as observed from the $\textrm{ER}\approx0.57$) for the externally applied magnetic field amplitudes $B=\pm3$~T. Since diffracted transmission is switched from $(0,0)$ to $(\pm1,0)$ or $(0,\pm1)$, the corresponding transmitted wavefronts will be deflected from $\theta_{0,0}=0^{\circ}$ (see Eq.~\eqref{eqgrating}) to $\theta_{\pm1,0}=\theta_{0,\pm1}=\pm45^{\circ}$ by the application of $B$. Indeed, numerical results for the gain (far-field profile) are shown in Figure~\ref{fig4}. For eye-guide, we illustrate the incident direction, cartesian axes, and directions of \textbf{B} in Figure~\ref{fig4}a). Results for $B$=0 and $B=\pm3$~T along the $\hat{\textbf{x}}$ and $\hat{\textbf{y}}$ directions are shown in Figure~\ref{fig4}b)-f). These results can be used for improved sensing~\cite{Wang2023} and THz beamforming,~\cite{tan2023terahertz,you2023mechanically} where the metasurface can be placed over a transmitter antenna to allow magnetic tuning of the phase array. Moreover, our idea can also find applications in the future Tera-WiFi concept (according to the recent IEEE Standardization 802.15.3d-2017~\cite{IEEE802.15.3d-2017}) for active THz links that communicate a fixed transmission system with at least 5 receiving antennas (see results in Figure~\ref{fig4}).

%========FIGURE 7============
\begin{figure}[!b]
\centerline{\includegraphics[width=0.9\columnwidth]{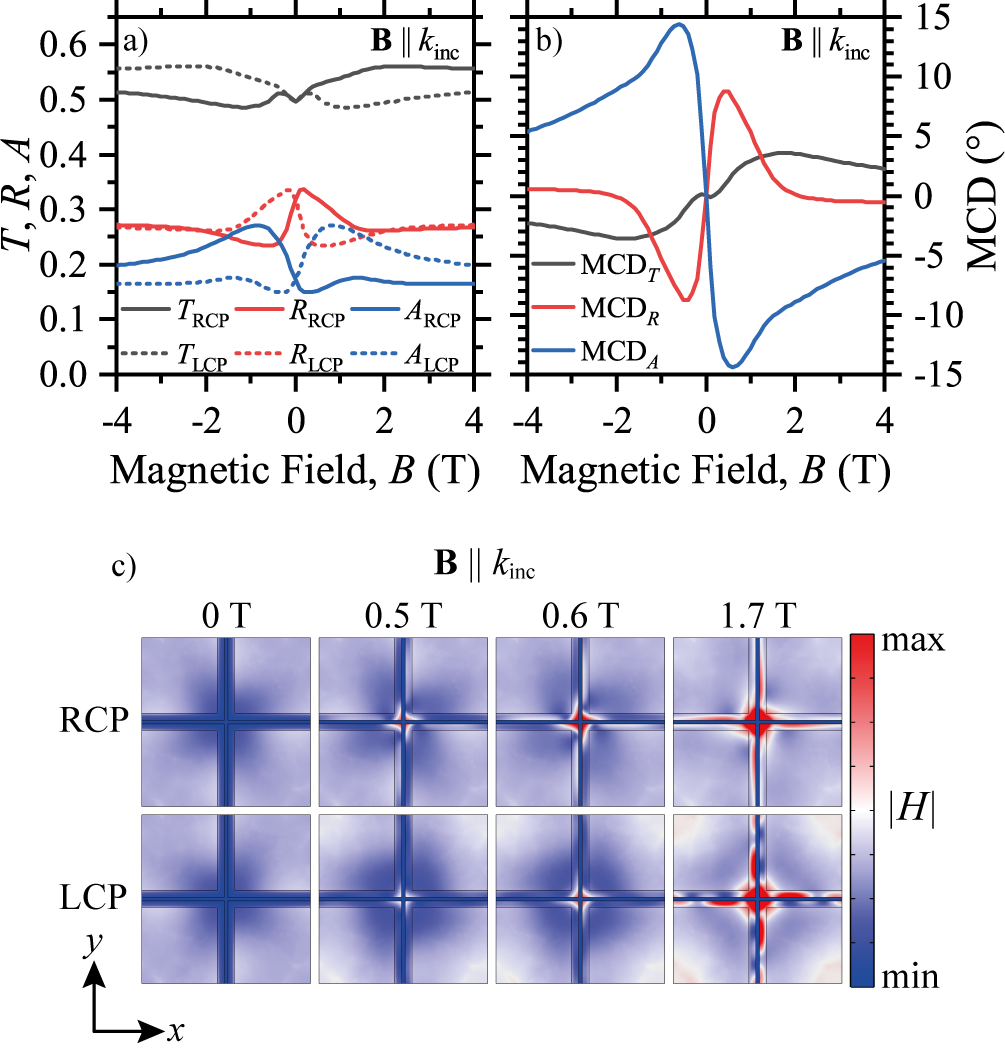}}
\caption{a) Total transmittance ($T$), reflectance ($R$) and absorptance ($A$) as function of $\textbf{M} \parallel \textbf{\^{z}}$ for RCP (solid curves) and LCP (dotted curves) at $f=300$~GHz. Magnetically induced differences in the results for RCP and LCP fields resulted in b) MCD for $T$ (black), $R$ (red) and $A$ (blue). c) Normalized near-field profiles (along the $xy$-plane) for the resonant $H$-field, as function of the applied magnetic field amplitude, where magnetically induced chirality (through the Faraday effect) is observed.}
\label{fig7}
\end{figure}
%==========================

As mentioned before, the magnetically switchable hyperbolic feature of the InSb material is responsible for the tunable THz beamforming in Figure~\ref{fig4}. To support this claim, we first plot the near-field profiles for RCP and LCP incident wavefronts in Figure~\ref{fig5}a)-c), where numerical results are comparatively shown for $T_{0,-1},T_{0,0},$ and $T_{0,1}$ with $B=-3~\text{T},B=0~\text{T},$ and $B=+3~\text{T}$, respectively. For simplicity, calculations are only shown for $\textbf{B}\parallel \textbf{\^{x}}$, since symmetric results are obtained for $\textbf{B}\parallel \textbf{\^{y}}$. It can be seen from these figures that an asymmetric dipole resonance is excited for $B=\pm3$~T. The latter is a HRI dielectric resonance occurring only along the $\textbf{\^{y}}$-axis, where $\varepsilon_{\rm \perp}>0$ (perpendicular to $\textbf{B}$), whereas the electromagnetic field is mostly expelled from the InSb slabs along the direction where $\varepsilon_{\rm \parallel}<0$, as noticed from Figure~\ref{fig5}c). Indeed, for $\textbf{B}=0$, no field is found within the building components of InSb, since both $\varepsilon_{\rm \parallel}$ and $\varepsilon_{\rm \perp}$ are simultaneously negative.

Having shown the dipole excitation associated with the magnetically switchable hyperbolic behavior of the InSb material, we now qualitatively explain the magnetically active beamforming through the dipole theory. Figure~\ref{fig6}a) illustrates (using LCP under $\textbf{B}=+3~\text{T}\parallel \textbf{\^{x}}$) how the resonance in the unit cell can be viewed, in a simplified form, as two nearby MO dipoles (spheres in the upper side). The external magnetic field (\textbf{B}) is applied parallel to the axis ($\textbf{\^{x}}$) along which the MO dipoles are placed. Figures~\ref{fig6}b)-c) show the scattered electromagnetic energy along the $xz$ and $yz$ planes, respectively, from where it can be clearly noticed that magnetically tuned beam tilting is only observed along the plane perpendicular to $\textbf{B}$. This tilting occurs due to the magnetically induced dipole moment along the $z$-axis, as it was demonstrated for a single dipole in Ref.~[\!\!\citenum{Carvalho2023a}]. Although not shown here, we must mention that the number of scattered lobes in the pattern of Figure~\ref{fig6}c) depends on the distance between the dipoles. We used here a relatively small distance to illustrate the tilting with only one lobe, whereas additional small lobes can be seen for larger distances. The latter becomes interesting if we see that the patterns in Figure~\ref{fig4} show some minor lobes in several different directions. Nevertheless, a direct comparison cannot be made since the metasurface also has diffractive effects, not included in the qualitative dipole approximation for the unit cell.

Figure~\ref{fig7} shows that the proposed metasurface can also exhibit magnetic circular dichroism (MCD) activity, which occurs when $\textbf{B}$ is applied parallel to the direction of propagation $\textbf{k}_{\rm inc}$ ($\textbf{\^{z}}$-axis in this case). Results are shown for the transmittance, reflectance and absorbance with solid and dashed lines for RCP and LCP incident wavefronts, respectively, in Figure~\ref{fig7}a). MCD can be defined in transmission, reflection and absorption using the corresponding amplitudes as\cite{mohammadi2018nanophotonic}
%==================================
\begin{equation}\label{eqMCD}
    \textrm{MCD} = \tan^{-1}\left(\frac{I_{\rm RCP}-I_{\rm LCP}}{I_{\rm RCP}+I_{\rm LCP}}\right),
\end{equation}
%==================================
in degrees ($^{\circ}$), where $I_{\rm RCP}$ and $I_{\rm LCP}$ are either the transmittance, reflectance, or absorbance amplitudes for the RCP and LCP wavefronts, respectively, with numerical results shown in Figure~\ref{fig7}b). Interestingly, since $\textbf{B}\parallel +\textbf{\^{z}}$, HRI dielectric resonances can be excited in the InSb slabs along the $xy$-plane, as corroborated in Figure~\ref{fig7}c), introducing magnetically tunable chirality in the metasurface.

%========================================================
\section{Conclusion}

We explored the magnetically switchable hyperbolic dispersion of the InSb material, combined with diffractive surfaces and magnetic tilting of scattered fields, to demonstrate a concept for dynamic terahertz beamforming. In particular, we use the conditions to transmit only the diffracted modes of $0$-th and $1$-st order, whose diffraction angles can be actively manipulated by the direction and sense of the applied magnetic field. The latter was achieved through the numerical design of the metasurface unit cell. Full-wave numerical results (using the finite element method) were found in qualitative agreement with semi-analytical results from the generalized dipole theory. A proof of concept was conducted using the operating frequency $f=300$~GHz, which is expected to provide efficient THz indoor wireless communications for future Tera-WiFi networks (according to the IEEE Standardization 802.15.3d-2017). Though results are shown for magnetic field amplitudes of $\pm3$~T, at temperatures of 200~K, it should be mention that recent experimental demonstrations of high-performance THz isolators (with InSb at room temperature) were made using field amplitudes of up to 0.4~T (which can be supplied by small permanent magnets).\cite{Lin2018,tan2023terahertz} Since magnetic field effects can be manipulated faster than temperature effects, our idea provides a new way to develop more rapidly reconfiguring THz metasurfaces.

%=================================
\acknowledgments
This work was partially supported by RNP, with resources from MCTIC, Grant No.01245.020548/2021-07, under the Brazil 6G project of the Radiocommunication Reference Center (Centro de Referência em Radiocomunicações - CRR) of the National Institute of Telecommunications (Instituto Nacional de Telecomunicações - Inatel), Brazil, and by Huawei, under the project Advanced Academic Education in Telecommunications Networks and Systems, contract No PPA6001BRA23032110257684. We also acknowledge financial support from the Brazilian agencies National Council for Scientific and Technological Development-CNPq (152370/2022-6, 314671/2021-8) and FAPESP (2021/06946-0).

%===========================================================
\bibliography{references}% Produces the bibliography via BibTeX.

\end{document}